\newif\ifsubmode
\submodetrue

\ifsubmode
  \documentclass[12pt,preprint]{aastex}
  \received{}
  \revised{}
  \accepted{}
\else
  \documentclass[12pt,preprint]{emulateapj}

\fi

\newcommand{\figonecap}{Comparison of axis ratios between the $V_{606}$
and $i_{775}$ bands for different magnitude and size cuts.  The
$\sigma_V$ and $z_{850}$ are the size of a galaxy in the $V_{606}$ band
and its magnitude in the $z_{850}$ band respectively.}

\newcommand{\figtwocap}{Comparison of ellipticity components in the
$V_{606}$, $i_{775}$, $z_{850}$ bands for large ($\sigma_V > 2.5$ pix)
and faint ($24 < z_{850} < 26)$ galaxies.  The $\sigma_V$ and $z_{850}$
are the size of a galaxy in the $V_{606}$ band and its magnitude in the
$z_{850}$ band, respectively.}

\newcommand{\figthreecap}{Comparison of ellipticity components among the
first three epochs in the $z_{850}$ band for large ($\sigma_z > 3$ pix)
galaxies.  The $\sigma_z$ and $z_{850}$ are the size of a galaxy and its
magnitude in the $z_{850}$ band.}

\newenvironment{inlinefigure}{
\def\@captype{figure}
\noindent\begin{minipage}{0.999\linewidth}\begin{center}}
{\end{center}\end{minipage}\smallskip}

\begin{document}
 
\title{Optimal Galaxy Shape Measurements for Weak Lensing
Applications Using the Hubble Space Telescope Advanced Camera
for Surveys}
 
\author{Yousin Park\altaffilmark{1,2}, Stefano Casertano\altaffilmark{2} \& Henry C. Ferguson\altaffilmark{1,2}}
\email{ypark, stefano, ferguson@stsci.edu}
 
\altaffiltext{1}{Department of Physics and Astronomy, The Johns Hopkins University, Baltimore, MD
21218}
\altaffiltext{2}{Space Telescope Science Institute, Baltimore, MD 21218}
 
\begin{abstract}

We present three-epoch multiband ($V_{606}$, $i_{775}$, $z_{850}$)
measurements of galaxy shapes using the ``polar shapelet'' or
Laguerre-expansions method with the Hubble Space Telescope ($HST$)
Advanced Camera for Surveys (ACS) data, obtained as part of the {\it
Great Observatories Origin Deep Survey} (GOODS).  We take advantage of
the unique features of the GOODS/ACS Fields to test the reliability of
this relatively new method of galaxy shape measurement for weak lensing
analysis and to quantify the impact of the ACS Point Spread Function
(PSF) on $HST$/ACS data.  We estimate the bias introduced by the sharp PSF
of the ACS on shape measurement.  We show that the bias in the
tangential shear due to galaxy-galaxy lensing can be safely neglected
provided only well-resolved galaxies are used, while it would be 
comparable to the signal level (1--3\%) for cosmic
shear measurements. These results should of be general utility in 
planning and analyzing weak lensing measurements with $HST$/ACS data. 

\end{abstract}


\section{Introduction}

The success and promise of weak lensing measurements (see Bartelmann \&
Schneider 2001 for a recent review) has gradually demanded new,
higher-precision methods to measure galaxy shapes because the weak
lensing signal is subtle.  The definition of ``galaxy shape'' for weak
lensing purposes is to some extent arbitrary, as long as it enables: (1)
to describe small changes in galaxy shape in sensible mathematical
terms; (2) to measure galaxy shapes efficiently when implemented in
practice; (3) to quantify various effects including PSF smearing; and
(4) to minimize the uncertainty in the galaxy shape measurements. 
Recently, Bernstein \& Jarvis (2002, BJ) have proposed a new method for
galaxy shape measurement using Laguerre expansions.  A similar approach
has been independently suggested as ``shapelet'' methods by Refregier
(2003) and Refregier \& Bacon (2003), who call the Laguerre expansion
``polar shapelets.'' A Laguerre expansion is based on the eigenfunctions
of a general two-dimensional quantum harmonic oscillator, which is
characterized by the oscillator's equilibrium position, its strength in
two perpendicular directions and the orientation of its symmetry axis. 
The oscillator's strength and shape define the size, ellipticity and
orientation of the Gaussian kernel.  Each such set of eigenfunctions
constitutes an orthonormal complete basis set that can be used to
decompose a galaxy image; see BJ for more details. 

The shapelet method has several advantages over other similar methods
(notably the formalism by Kaiser, Squires \& Broadhurst 1995) which 
use Gaussian-weighted second moments for shape measurement.  The
shapelet method: (1) applies the Gaussian weight adaptively according to
the ellipticities of objects, which leads to better shape measurements
for highly elliptical objects; (2) is potentially more efficient, since
most operations can be performed by manipulating the shapelet
coefficients of images, which can be expressed as a Hermitian matrix,
instead of handling pixel data directly; and (3) allows better
corrections of various effects on galaxy shape measurements, since most
of those effects can be expressed as transformation matrices acting on
the coefficient matrix.  Since its introduction, the method has been
used for cosmological weak shear measurements with ground-based observations (e.g., Jarvis et al.\ 2002) and morphology studies (e.g., Massey et al.\ 2003a). 
We have independently implemented the method and successfully measured the tangential shear due to galaxy-galaxy lensing as a first application (see Casertano et al.\ 2003). 

Weak lensing measurements impose demanding requirements on observations. 
Ideally one would like to have deep observations over a wide area with
an extremely small PSF.  To control systematics, it is best if the
observations are taken at multiple orientations, with galaxies appearing
at different places on the detector.  Multiband measurements have been
adopted along with null tests to secure confidence in detection of weak
lensing signals (e.g., McKay et al.\ 2001).  The GOODS/ACS observations
are perhaps the best data set yet obtained for these purposes.  The
general advantages of space-based observations for weak lensing are
discussed by Rhodes et al.\ (2003) and Massey et al.\ (2003b).  The ACS
provides excellent sensitivity with a small PSF width.  Observations of
just a few orbits provide a high surface density of galaxies with
signal-to-noise ratios ($S/N$) sufficient for shape measurements.  Due
to the supernova search strategy (Giavalisco et al.\ 2003), the GOODS
observations were spaced into multiple epochs, separated by 40 to 50
days.  Each epoch was observed at a different orientation due to HST
observing constraints.  In this letter we compare galaxy shape
measurements from {\it independent} subsets of the GOODS data in order
to assess the sensitivity and robustness of the measurement technique and the quality of $HST$/ACS data for weak lensing studies. 

\section{Observations and Measurements}

The GOODS/ACS fields are imaged in the ACS F435W, F606W, F775W, and
F850LP filters (hereafter referred to as $B_{435}$, $V_{606}$, $i_{775}$,
and $z_{850}$).  The observations ultimately reach $S/N \sim 10$ limits
$27.2, 27.5, 26.8, 26.7$ AB mag for faint galaxies; for more details of
the GOODS/ACS Fields, see Giavalisco et al.\ (2003).  The work in this
letter is based on images of the first three epochs of the GOODS/ACS
Chandra Deep Field South (CDF-S), which covers approximately $10 \arcmin
\times 16 \arcmin$ at $0 \arcsec .05$ pix$^{-1}$.  Only the $V_{606}$,
$i_{775}$ and $z_{850}$ band images are used in this letter.  For each epoch,
the total exposures are approximately $1050$, $1050$, and $2100$
seconds, respectively.  The field orientations of two successive epochs
differ by $45\deg$. 

The {\bf SExtractor} software (Bertin \& Arnouts 1996) is used to detect
and catalog objects (predominantly galaxies with some stars) and create
masks for objects.  The objects are extracted into individual images
using the {\bf SExtractor} segmentation map.  Each object is centered in
a rectangular postage-stamp image with dimensions that are at least
twice the {\bf SExtractor} extent of the object.  The minimum size of
the images is $32 \times 32$ pixels.  We mask out surrounding objects in
each of these images.  For the weak-lensing measurements, we discard
galaxies larger than $100 \times 100$ pixels and stars because they are
almost all nearby objects that are irrelevant to lensing study.  Once
the postage-stamp images are prepared, we use the shapelet method to
measure galaxy shapes. 

The essence of the shapelet method is to approximate the intensity image
of a galaxy as a finite sum of Laguerre functions with kernel
properties---center, size, and shape---that match those of the image. 
The optimally matched kernel is the one in which the terms of the
Laguerre expansion satisfy the centroid, roundness, and significance
conditions, as defined in BJ.  [Explicitly, these conditions mean that
the (1,0), (1,1), and (2,0) terms of the Laguerre expansion vanish.] The
size and shape of the galaxy are then {\it defined} to be the size and
shape of the optimally matched kernel; BJ show that this definition
obeys the correct transformation rules in the presence of translation,
dilation, and shear, as may be due to weak lensing, and furthermore the
definition is {\it optimal} in the signal-to-noise sense.  In addition,
the method provides a way to correct for PSF effects via a deconvolution
expressed as an operation on the coefficients of the Laguerre expansion. 
We implemented the shapelet method within the Interactive Data Language
({\bf IDL}).  Our current implementation successfully measures the
position, size ($\sigma$), axial ratio ($b/a$) and position angle (PA)
for $ \sim 95\% $ of the galaxies brighter than $ z_{850} = 24 $ and $
\sim 92\% $ of all galaxies down to $ z_{850} = 28 $; for the remaining
sources, the method fails to converge.  When including PSF corrections,
the failure rates become $1.5-2$ times higher. 

\section{Measurement Results}

The ellipticity, shear or shape of a galaxy can be represented as
 \begin{equation}
 \vec{e} \equiv (e_1, e_2) \equiv (|\vec{e}| \cos 2 \phi, |\vec{e}|
    \sin 2 \phi ),
 \end{equation}
 where $ \phi $ is the position angle of the galaxy major 
axis and $|\vec{e}| = (a^2 - b^2) / (a^2 + b^2)$; see Miralda-Escud\'{e}
(1991).  Weak lensing distorts the shape slightly: $\vec{e}_{\rm dis} = \vec{\delta}_{\rm WL} \oplus \vec{e}_{\rm int}$, where $\vec{\delta}_{\rm WL}$ is the shear due to weak lensing and
$\oplus$ is the shear addition operator.  Here, $\vec{e}_{\rm dis},
\vec{e}_{\rm int}$ are the distorted and intrinsic shapes of the galaxy, respectively.
The {\it shape noise}, variation in intrinsic galaxy shape, is large:
$\sigma_{\rm int} = \sqrt{\langle |\vec{e}_{\rm int}|^2 \rangle} \approx
0.62$, which is substantially larger than weak lensing signal
$|\vec{\delta}_{\rm WL}| \ll 1$.  This makes weak lensing measurements
challenging.  To make matters worse, noise (predominantly due to sky
background) and systematics (predominantly due to PSF) introduce {\it
measurement errors} in shape: $\vec{e}_{\rm obs} = \vec{\delta}_{\rm PSF} \oplus \vec{\delta}_{\rm SB}
     \oplus \vec{e}_{\rm dis}$, where $\vec{e}_{\rm obs}$ are the shape of observed galaxy and
$\vec{\delta}_{\rm PSF}, \vec{\delta}_{\rm SB}$ are measurement
errors, which can be expressed as equivalent shears, due to PSF and
sky background respectively. These errors demand more complicated
analysis for lensing signal. For the analysis, it is essential to
have good knowledge on $\vec{\delta}_{\rm SB}$ (mostly stochastic)
and $\vec{\delta}_{\rm PSF}$ (mostly systematic). The multiepoch and
multiband measurements provide an ideal way to assess their tendencies.

The results presented here are based on the images without PSF
correction.  The ACS PSF is described by Ford et al.\ (2003) and
Pavlovsky et al.\ (2002); its full width at half-maximum in the GOODS
images ranges from $0\farcs12$ to $0\farcs 15$ depending on filter.  The
PSF is close to isotropic after removing the geometric distortion: the
shapelet decomposition of a composite PSF yields $\sigma, b / a, {\rm
PA} = (1.15 {\rm \; pix}, 0.985, 78\fd 1) $, $ (1.17 {\rm \; pix},
0.969, 59\fd 8)$, $ (1.25 {\rm \; pix}, 0.941, 83\fd 7)$ for the
$V_{606}, i_{775}, z_{850}$ bands respectively, with very small field
dependence.  The PSF is expected to affect the measured shapes in two
ways: by broadening the galaxy image, making the galaxy look larger and
rounder; and by imparting a preferential angle to the apparent
orientation, a consequence of the slight anisotropy of the PSF itself. 
The former affects primarily the estimate of the susceptivity of galaxy
images to gravitational distortion and will not introduce a spurious
lensing signal or change it to first order.  On the other hand, a
possible bias on the observed orientation could add directly onto the
measured signal, and therefore can be potentially problematic.  In order
to estimate the possible bias introduced by the PSF, we compare $\langle
\vec{e} \rangle$, the average galaxy shape over a large number of
galaxies in different bands and in different epochs.  Any systematic PSF
dependence would manifest itself as a difference in the average values
of the ellipticity components.  Measurements in different epochs are
especially significant, since the PSF rotates between epochs. 

For the band-to-band comparison in shape measurement, we use a combined
three-epoch mosaic for each band; the image catalog is obtained from 
the $z_{850}$ band image and contains $\sim 25000$ objects.  The galaxy masks
obtained from the $ z_{850} $ band image are used for the other bands as
well.  We limit our comparison to galaxies with $ 22 < z_{850} < 26 $ 
for which the shapelet measurement succeeds in all three bands.
We divide the galaxies into bright ($22 < z_{850} < 24$) and faint ($24
< z_{850} < 26$), and into large ($\sigma_V > 2.5$ pix) and small
($\sigma_V < 2.5$ pix), where $\sigma_V$ is the Gaussian size of the
galaxy in the $V_{606}$ band.  Figure~1 presents a qualitative comparison of axis ratios measured in the $V_{606}$ and $i_{775}$ bands.  The axis ratios 
are generally in good agreement for all magnitude and size cuts.  Small
galaxies appear to be rounder than larger ones regardless of brightness,
presumably as a consequence of PSF dilution.  A closer look indicates
that galaxies look slightly rounder in the $ i_{775} $ band than in the $V_{606} $ band,
in agreement with the slightly larger PSF in the former band. 
Similarly, the position angle (not shown) compares well on average
between the bands, with no net difference and a dispersion that 
increases for fainter, smaller, and rounder galaxies, as might be expected
since the position angle is more difficult to measure for such galaxies.
These qualitative comparisons show that the shapelet method works 
generally well on real galaxy images.

However, more important for weak lensing studies is a {\it quantitative}
comparison of the ellipticity components measured on galaxies as a
function of band and epoch.  The ellipticity components $ e_1, e_2 $
(defined in Eq.~1) are the quantities used in weak lensing; systematic
differences, as could be due, e.g., to PSF or to pixellation effects,
would affect directly the ability to carry out weak lensing
measurements.  Figure~2 shows the band-to-band comparison of the
ellipticity components for the large ($\sigma_V > 2.5 $), faint ($ 24 <
z_{850} < 26 $) galaxies in Figure~1. We average the difference in
ellipticity components for the galaxies in Figure~2 , and find $\langle
\vec{e}(i_{775}) \rangle - \langle \vec{e}(V_{606}) \rangle = (0.0003
\pm 0.0036, 0.0054 \pm 0.0035)$, $\langle \vec{e}(z_{850}) \rangle -
\langle \vec{e}(V_{606}) \rangle = (-0.0004 \pm 0.0039, -0.0144 \pm
0.0039)$.  Thus, the net difference between ellipticities measured in the $
V_{606} $ and $ i_{775} $ bands is $\sim 0.5\% $; between the $ V_{606} $ and
$z_{850} $ bands is $ \sim 1.4\% $.  These net differences are almost entirely
due to the PSF, and are more significant in the $z_{850}$ band because of the larger, less round PSF. 

\ifsubmode
\else
\begin{inlinefigure}
\begin{center}
\resizebox{\textwidth}{!}{\includegraphics{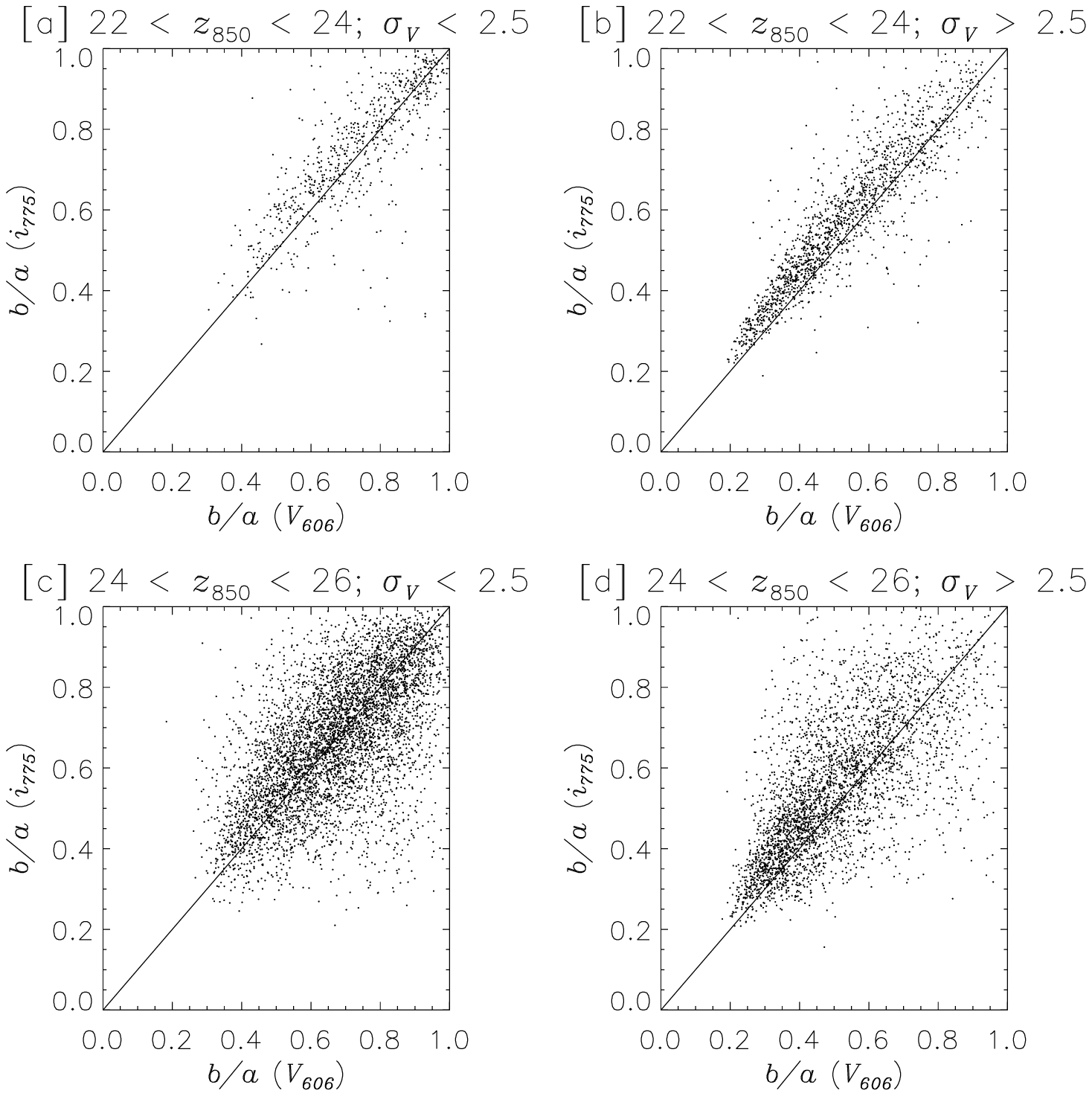}}
\figcaption{\figonecap}
\end{center}
\end{inlinefigure}
\fi

\ifsubmode
\else
\begin{inlinefigure}
\begin{center}
\resizebox{\textwidth}{!}{\includegraphics{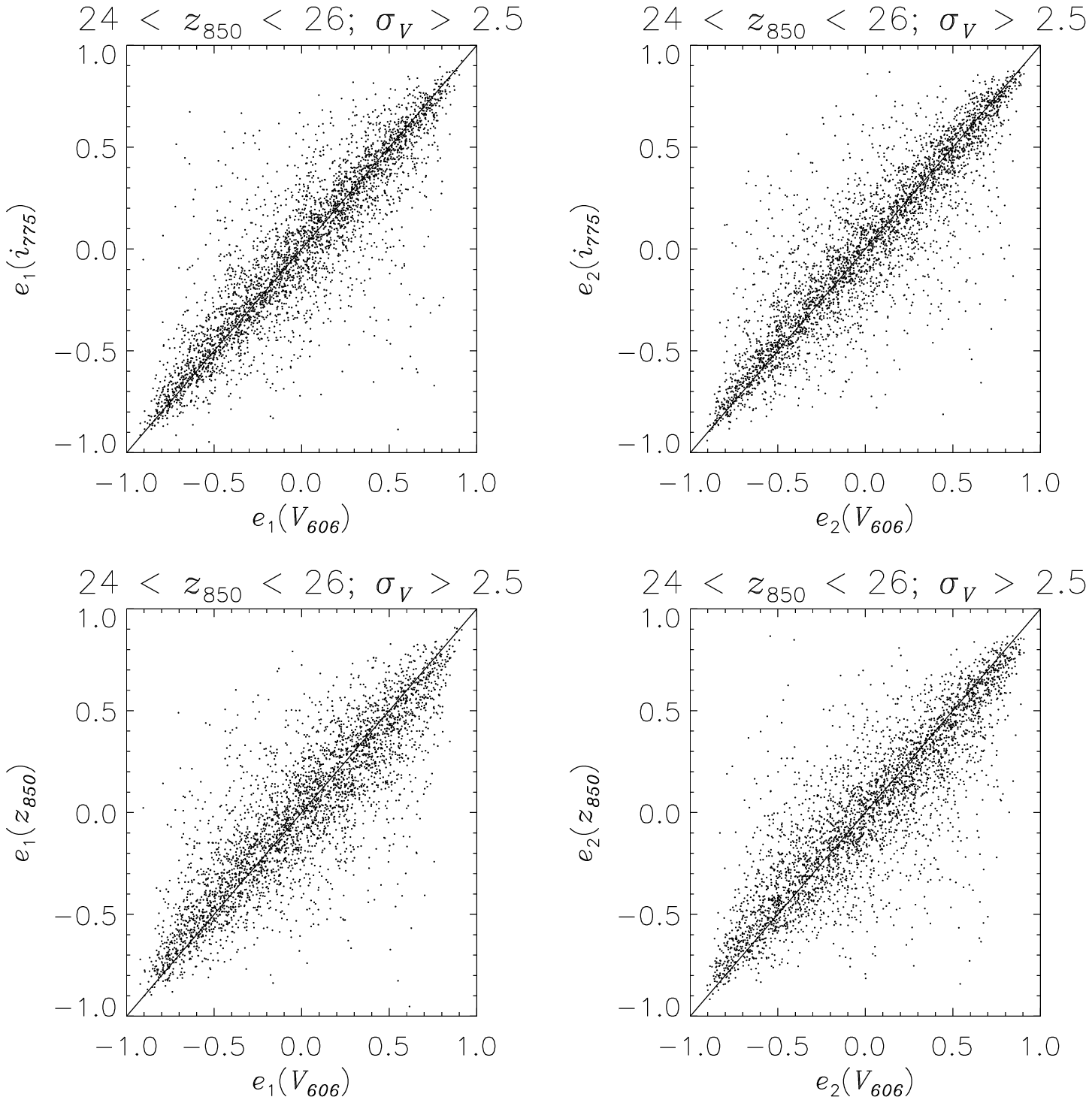}}
\figcaption{\figtwocap}
\end{center}
\end{inlinefigure}
\fi

The most direct test of the stability of shape measurements can be
obtained by comparing measurements in the same band but at different
epochs.  We can only carry out such a test for the $ z_{850} $ band 
images, since in the
other bands not enough images were taken at each epoch for a good
rejection of cosmic rays.  At 
different epochs, the GOODS/ACS fields put galaxies on different pixel
positions, and the detector rotates by $ 45 \deg $ between consecutive epochs.
The images have slightly different sky backgrounds, and the {\bf
SExtractor} segmentation maps for the galaxies differ from
epoch to epoch due to the effects of noise.  PSF anisotropies and
variations over time will distort galaxy shapes differently in different
epochs, as will the geometric distortion of the camera and the
appearance of hot pixels, unrejected cosmic rays, and so on.  
As each single-epoch image is shallower than the combined image, the
catalog for the single-epoch images is about half the size of 
the three-epoch catalog. The comparison of ellipticity components between epochs is shown in
Figure~3, limited to galaxies with size $\sigma_z > 3$ pix.  There
is good agreement between the ellipticity components measured in
different epochs.  The scatter between each pair of epochs is
comparable, suggesting that the variation in each object's measurement
is primarily due to sky background.  Since the PSF is
not perfectly round, it introduces bias.  For the galaxies in Figure~3, we obtain $\langle\vec{e}({\rm Epoch2}) \rangle - \langle \vec{e}({\rm Epoch1}) \rangle = (-0.0175 \pm 0.0073, 0.0158 \pm 0.0074)$ and $\langle \vec{e}({\rm
Epoch3}) \rangle - \langle \vec{e}({\rm Epoch2}) \rangle = (-0.0153 \pm
0.0070, 0.0116 \pm 0.0070)$, which indicates a systematic change of
$\sim 2\%$ in $\langle \vec{e} \rangle$ between epochs.  This change is
consistent with the change in the PSF on the sky---due to the rotation
of the field of view between epochs---and implies that, even for ACS
data, correction for PSF effects is necessary for cosmic shear
measurements (especially with the $z_{850}$ band images), where the signal is typically 1--3\%. 

The situation is different for galaxy-galaxy lensing shear measurements, which
are expressed in terms of the tangential ellipticity $ e_+ \equiv
|\vec{e}| \cos2 \theta $, where $\theta$ is the angle between the major
axis of a source and the line connecting its center to the center of a
lensing galaxy.  We find no statistically significant correlation
between ellipticity change and position in the detector, implying that
the effect of the PSF is constant over the field of view.  Therefore the
PSF-induced difference in ellipticity can be expected to average out,
due to the random placement of lens-source pairs.  Indeed, we find that
the epoch-to-epoch difference in $ \langle e_+ \rangle $ for the
lens-source population used by Casertano et al.~(2003) is less than
0.01\%, consistent with the noise level and much smaller than the 
signal they detect, which drops to $ \sim 0.2\% $ at large separations.
We find similarly that the difference in $ \langle e_+ \rangle $ 
between bands is less than 0.01\%. 

In order to quantify the effect of PSF on shear measurements, we convolve
the galaxy images with the composite ACS PSF in the $V_{606}$ band, repeat the shape measurements
and compare the results with the original. The effect of the PSF on the shape of a well-resolved galaxy
($\sigma_{\rm galaxy} \gtrsim 2 \sigma_{\rm PSF}$) is expected to be
small, scaling roughly as $(\sigma_{\rm PSF} / \sigma_{\rm galaxy})^2$. 
 A round PSF will make a regular galaxy larger and rounder, but will not
change its orientation.  Indeed, we find that PSF convolution makes
galaxies larger and slightly rounder, with a change that decreases with
increasing galaxy size.  Due to the anisotropy of the PSF, the
convolution does induce some error in $e_1, e_2$, but the scatter is
substantially smaller than seen in the band-to-band or epoch-to-epoch
comparisons.  For the galaxies with $24 < z_{850} < 26.5 $ and $
\sigma_V > 2.5$, we find the difference in $\langle \vec{e} \rangle$
is $\delta \vec{e} = (-0.0044 \pm 0.0020, -0.0004 \pm 0.0020)$.  This
bias is associated with the slight anisotropy of the PSF; if uncorrected,
it would induce an ellipticity correlation at the level of $\sim \delta
\vec{e} / 2 \approx 0.2\% $, small but not negligible in some cases of
cosmological weak shear measurement.  As for the tangential shear
measurements, the bias is less than $0.01 \%$ by the tangential
averaging, and therefore can be safely neglected. Therefore in ACS data, the PSF correction 
(at least for the $V_{606}$ band) is not essential for the shear measurement of
galaxy-galaxy lensing provided only well-resolved galaxies are used. 

\ifsubmode
\else
\begin{inlinefigure}
\begin{center}
\resizebox{\textwidth}{!}{\includegraphics{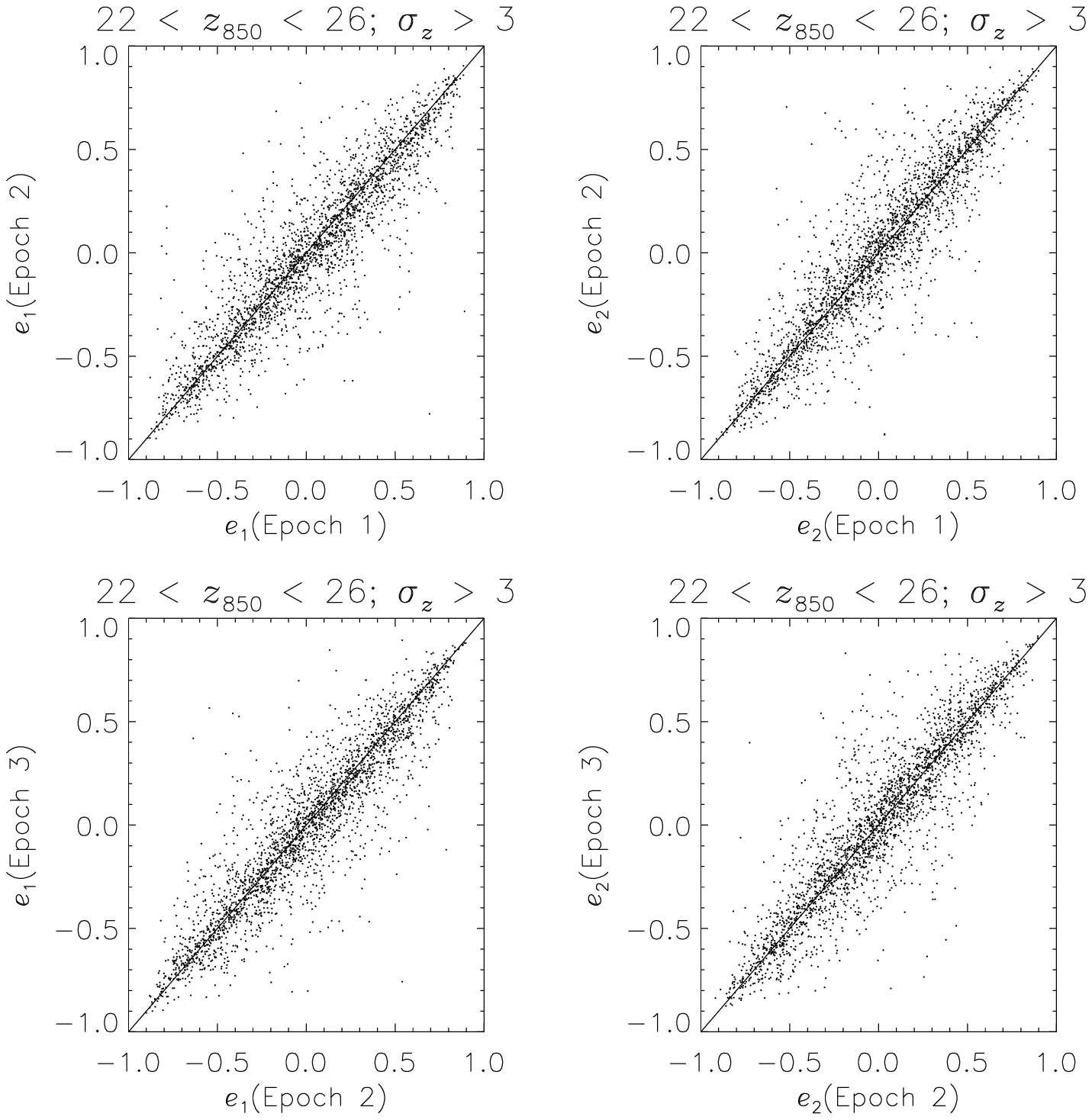}}
\figcaption{\figthreecap}
\end{center}
\end{inlinefigure}
\fi

Finally, we have carried out a blind test to determine whether we could
recover a known shear. We distort a part of the $z_{850}$ band image with $2\%$ shear in various directions, and compare the shapes of galaxies in the sheared images with original shapes. We can recover each component of the induced $2\%$ shears with relative errors of $\lesssim 10\%$ by comparing $\sim 3000$ galaxies in the images used in the test. 

\section{Discussion}

We have presented shape measurements of galaxies in $HST$/ACS data using
the Laguerre-expansion method.  Though more tests
of the PSF deconvolution step remain, the results here indicate that the
impact of the PSF on galaxy shape measurements with these data are not
the limiting factor for tangential shear measurement due to the superb
resolution of $HST$/ACS, provided the galaxies are well resolved.  The
level of PSF-induced bias appear to be acceptable for
tangential shear measurements, provided the background source galaxies
are limited to $z_{850} < 26.5$ and $\sigma_V > 2.5$ pix in our case.  Our
method and result can be applied to different ACS data sets for quick
tangential shear measurement. 

Though the method proposed by BJ is simple and elegant, we have found
that it needs some modification and adjustment in practice.  We have
done extensive tests with simulated galaxy images to assess the impacts
of such effects as different pixel scales and choices of drizzling
parameters.  Improvement of current implementation is also underway to
reduce the failure rate.  The detailed implementation procedures and
test results will be presented in a future article.  In addition to the
measurement of tangential shear due to galaxy-galaxy lensing, we intend
to apply this technique to cosmological weak shear measurement, where
the impact of PSF effect is expected to be larger.  The depth of the
GOODS/ACS Fields is ideal for such measurement, although the relatively
small field of view (by ground-based standards) limits the applicability
to relatively small spatial scales.  The deepest $V_{606}$ band images
provide $\sim 20$ background sources per square arcmin with $z > 2$ and
$S/N > 15$, which we anticipate will be be sufficient for measuring the
trend of the weak shear amplitude on arcmin scales with redshift. 
Beyond measuring galaxy shapes for weak lensing, the the shapelet method
is a potentially valuable tool for studying the morphological evolution
of galaxies. 

\acknowledgments

We are grateful to GOODS Team, especially Tomas Dahlen and Swara
Ravindranath for their assistance in generating special catalogs for the
tests described in this letter.  We thank Norman Grogin for his help
with the blind test for image distortion.  Support for this work was
provided by NASA through grant GO09583.01-96A from the Space Telescope
Science Institute, which is operated by the Association of Universities
for Research in Astronomy, under NASA contract NAS5-26555. 

\ifsubmode
\newpage
\fi

\bibliographystyle{apj}

\ifsubmode

\begin{figure}
\figurenum{1}
\plotone{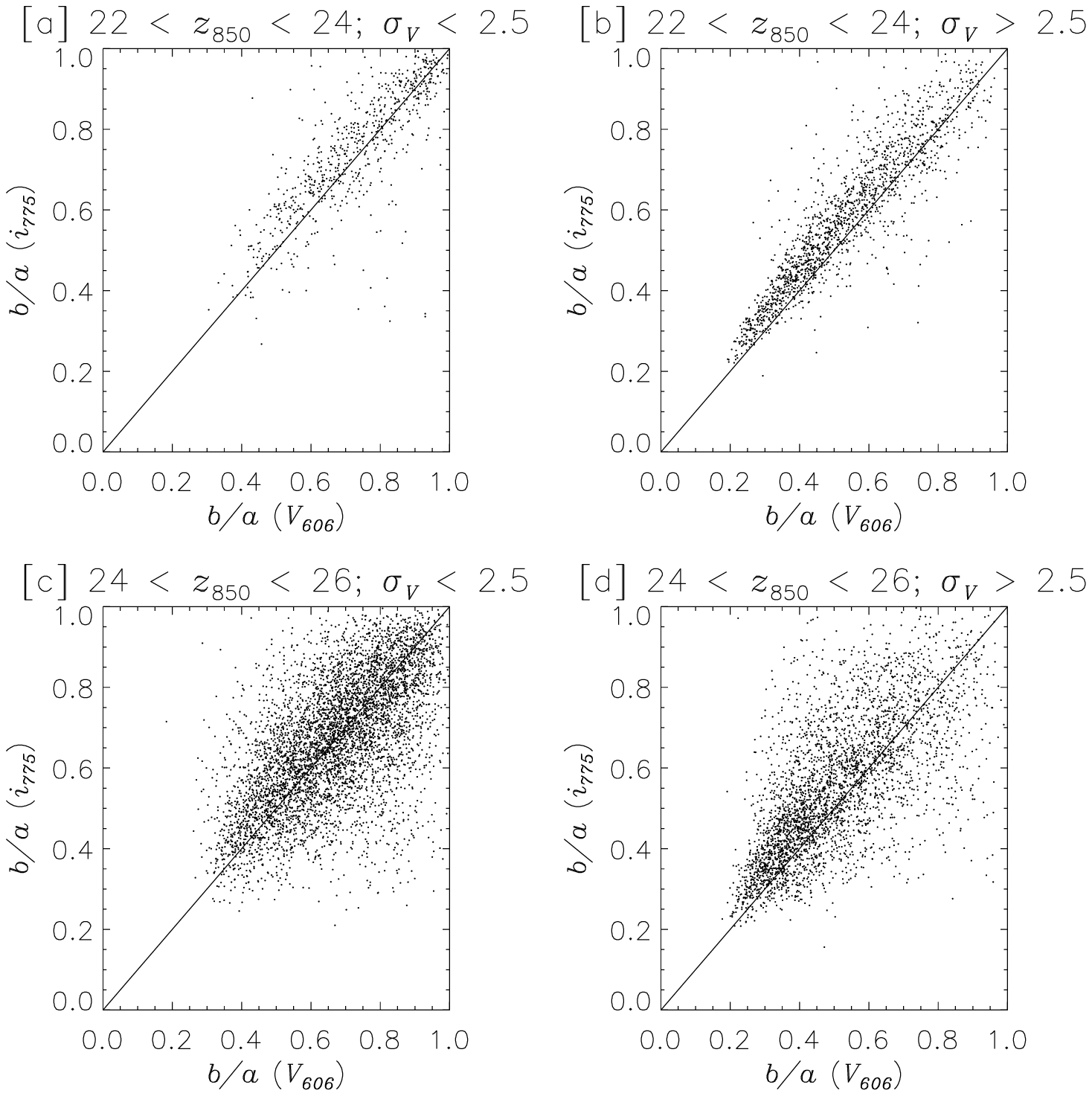}
\caption{ \figonecap } 
\end{figure}

\begin{figure}
\figurenum{2}
\plotone{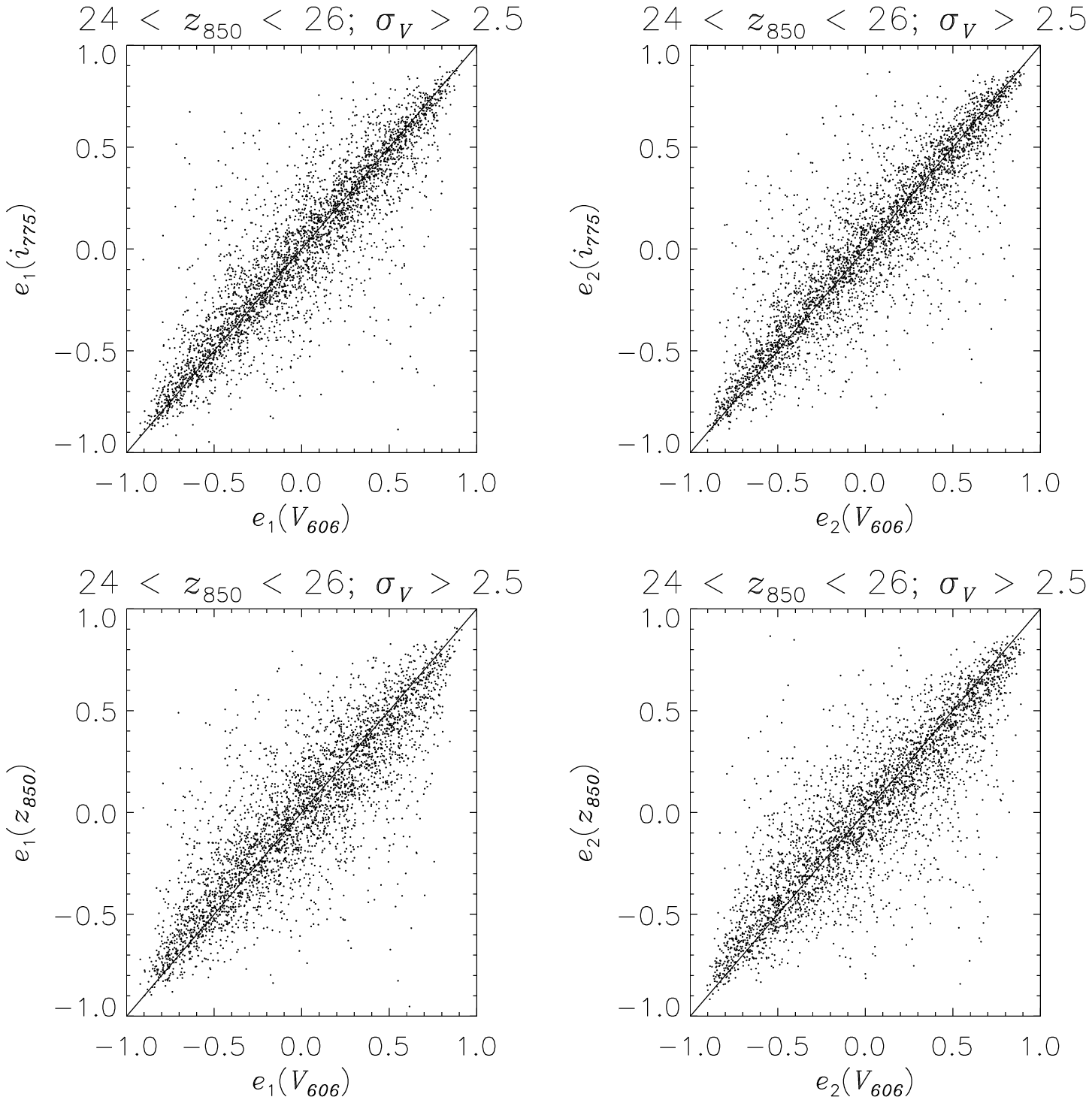}
\caption{ \figtwocap }
\end{figure}

\begin{figure}
\figurenum{3}
\plotone{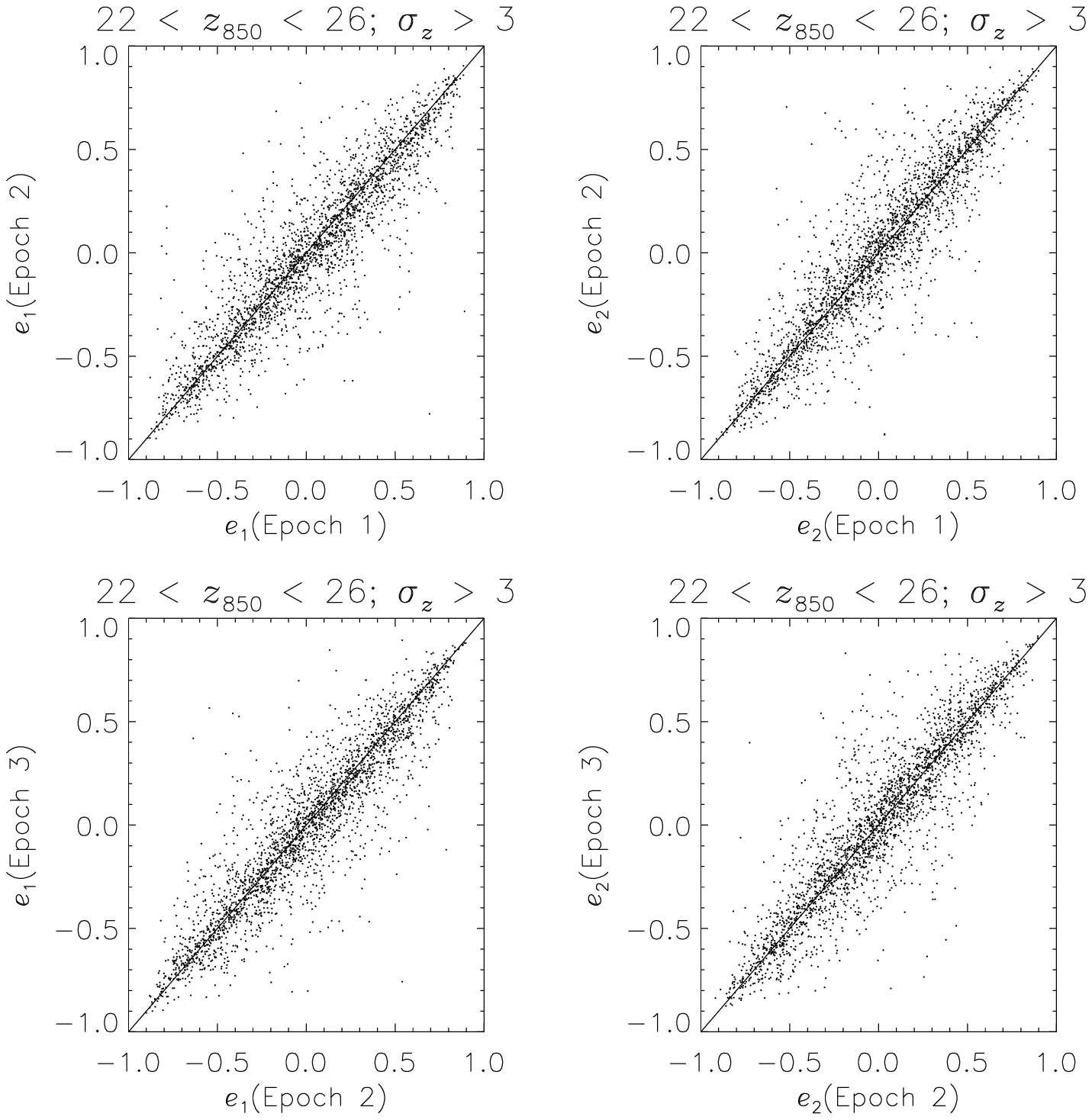}
\caption{\figthreecap}
\end{figure}

\fi

\end{document}